\documentstyle[12pt,fleqn]{article}
\begin{document}
\begin{titlepage}
\flushright{DUKE-TH-92-41}
\vspace{10cm}
\center{\large Lyapunov Exponent of Classical SU(3) Gauge Theory}
\vspace{0.8cm}
\center{Chengqian Gong}
\center{\em	Physics Department, P.O.Box 90305, Duke University}
\center{\em Durham, NC 27708-0305, U.S.A.}
\vspace{1.5cm}
\begin{abstract}
The classical SU(3) gauge theory is shown to be
deterministic chaotic. Its largest Lyapunov exponent is dertermined,
from which a short time scale of thermalization of a pure gluon system
is estimated.
The connection to gluon damping rate is discussed.
\end{abstract}
\end{titlepage}

\bigskip
\bigskip
\newpage

In Ref.[1], a classical lattice gauge method
was introduced showing that SU(2) gauge theory is
chaotic in the semi-classical limit. The largest
Lyapunov exponent was obtained. In this Letter,
we use the same method to study SU(3) gauge theory.
While the study of SU(2) theory gives some hints
on properties of a general non-abelian gauge theory,
the present study of SU(3) will
give us some direct and specific
information about a highly excited gluon system.
The reason why such a semi-classical approach is
relevant here
is that in a highly excited system the
contribution of quantum fluctuations to some observables (for
example, Debye screening mass, gluon damping rate, we will
explain this briefly in the end of this Letter) is
negligible compared to those from thermal
excitations. So at least for some observables, classical
calculations can give correct answers. Here
we want to use this method to study entropy
production and thermalization in a highly excited
gluon system. This process is relevant
in the study of the early universe \cite{olive} and of
relativistic heavy ion collisons \cite{muller1,mclerran,hwa},
where the phase transition of nuclear matter
to quark-gluon plasma is expected to occur. The
proposed signatures of this transition depend critically on
the time scale of the pre-equilibrium thermalization process.
Studies \cite{thoma,shuryak}
show that the time scale of this process is short
and the gluon sector thermalizes much faster
compared to the quark sector($\tau_{q}=3\tau_{g}$).
This later fact makes it plausible to consider the thermalization of
the gluon sub-system separately. But
up to now, this process is still not completely understood
because a fully non-perturbative quantum mechnical calculation
is beyond our ability. On the other hand, our
method not only leads to an exact and non-perturbative solution
to this problem
(in the classical limit), but
also helps to explain
the reason of a fast thermalization process
in the classical language. That is
the chaotic nature of the non-abelian gauge theory.
We hope that our results will be helpful to a full
quantum treatment in the future.

This Letter is organized as follows. First we
will outline the method of a classical Hamiltonian
lattice simulation. Some technical points will be explained.
Then we will proceed to the results. An exponentially
increasing distance between two field configurations
is observed. From this behavior, an energy dependent Lyapunov
exponent is obtained which leads to an estimate of thermalization
time scale for the gluon system. Finally we will discuss
the relation between Lyapunov exponent and gluon damping rate.

The framework of a classcal lattice simulation in
Hamiltonian formalism of SU(2)
gauge fields can be found in Ref[1].
Here I want only to repeat some basic
points and to emphasize the differences
between SU(3) and SU(2) gauge fields.
The Hamiltonian for SU(3) gauge fields on a lattice \cite{kogut,chin} is
\begin{equation}
H=\frac{g^{2}}{a}(\sum_{l}\frac{1}{2} E^{a}_{l} E^{a}_{l}+\lambda
{\rm Re}
 \sum_{p}[1-\cos(\frac{1}{2} B_{p})]),
\end{equation}
where $g$ is the coupling constant, $\lambda=4/g^{4}$,
and $a$ is the lattice spacing.
In this calculation $\lambda=1.1185$ and $a=1$. The
particular choice is not important here because the
final result can be written in a scaling fashion(10).
The electric and magnetic fields
can be expressed in terms of the link variables
$U_{l}=\exp(-i\frac{1}{2}\lambda^{a}A^{a}_{l}),
$ which 
are elements of the group SU(3).
A set of classical
equations can be derived from (1):
\begin{eqnarray}
\dot{U_{l}} &=& i \sum_{a} V^{a}_{l} \lambda^{a} U_{l},\\
\dot{V^{a}_{l}}& = &\frac{1}{4 a^{2}} i
\sum_{p} {\mbox tr} \lambda^{a}(U_{p}-U^{\dagger}_{p}),
\end{eqnarray}
where
\begin{equation}
V^{a}_{l} = \frac{g^{2}}{2 a} E^{a}_{l},
\end{equation}
and $U_{p}$ is the product of all four link variables on an elementary
plaquette.
These equations define a trajectory \{$E^{a}_{l}(t),U_{l}(t)$\}
in the classical phase space
for each given initial condition
$\{E^{a}_{l}(0),U_{l}(0)\}$. In order to study the
exponential divergence of two trajectories we introduce the
following gauge-invariant distance between two trajectories 1,2:
\begin{equation}
D(1,2)=\frac{1}{2N_{p}}\sum_{p}\mid {\mbox tr} U_{1p} - {\mbox tr} U_{2p} \mid
{}.
\end{equation}
$D$ is propotional to the absolute local difference in the
magnetic energy of two different gauge fields.
$N_{p}$ is the total number of plaquettes: $N_{p}=3N^{3}$,
where $N$ is the lattice size. In the following calculation,
we use a $10^{3}$ lattice.

Initially we put all $E^{a}_{l}=0$ to avoid the presence of local static
color charges on the lattice.
Link variables are parametrized by eight
real angles $\alpha^{a}, a=1,...,8$ \cite{bronzan}.
Here there are two problems that do not exist in SU(2)
simulation.
First, in the case of SU(2) the 3 angles
are used throughout the simulation.
The group multiplication can be easily realized
in this representation.
But in the case of SU(3), if the 8 $\alpha$'s are directly used,
the group multiplication is difficult to realize.
We find that it is much faster to directly use the unitary
$3\times3$ matrices in the program.
Second, in SU(2) the initial magnetic energy put on the lattice
is controlled by a single angle $\omega$. Changing the
range of $\omega$ changes the energy. But we find in our
case, if we change the range of the angles in
order to change the energy, we
run the risk of restricting the initial configuration
to a small subspace of SU(3). So here
we use a different method to initialize the link variables.
We start from a random field configuration obtained by
arbitrarily selecting $\alpha's$ for each link.
Then we use a heat bath algorithm
\cite{pietarinen} to thermalize the
magnetic sub-system to temperature $T_{0}$. Now $T_{0}$
is the only parameter to control the total energy on the lattice.
A second configuration is chosen in the
close neighborhood of the first one ($D(0)$ is small). This can
be achieved by choosing slightly different $\alpha$'s (controlled
by a parameter $\delta$,$\mid\delta\mid<<1$)
for each corresponding link of the second configuration with
respect to that on the first one.

We numerically integrate the (5) and (6).
Using the measure in (8), we observe a similar behavior here
of divergence of two trajectories as for SU(2) in Ref[1].
In Fig.1.a, the evolution of $\ln(D(t))$ for $T_{0}$
is shown.
We see after several initial oscillations, $D(t)$
increases exponentially with $t$ and then saturates at large
$t$ due to compact nature of SU(3) group. We notice
in the exponential increasing region the fluctuation
is small for large $T_{0}$. This suggests the divergent
property is very similar in the entire phase space
except for some regions of small measure.  We identify
the slope of $\ln(D(t))$ in the linear region
as the largest Lyapunov exponent $h$.

We have also tried a different definition
of distance
\begin{equation}
D^{E}(t,1,2)=\frac{1}{2N_{p}} \sum_{l} \mid \sum_{a}
( (E_{1l}^{a})^{2}-(E_{2l}^{a})^{2} ) \mid,
\end{equation}
namely the sum of the absolute
value of local difference in the
electric energy. The rising
of $\ln(D^{E}(t))$ shown in Fig.1.b
is coincident with $ln(D(t))$
except for the initial oscillatory region.
This is somehow within our expectation
because the chaoticity of a system is an
intrinsic property and it shall not
depend on a particular metric we choose for it.

We have studied the wave-length dependence of
the trajectory divergence.
We define a local distance as
\begin{equation}
D(1,2,t,{\bf p})=\mid trU_{1p} - trU_{2p} \mid,
\end{equation}
which is a function of $t$ and $\bf p$. Here $\bf p$ denotes
the relative position of a particular
plaquette on the 3-dim lattice.
$D(1,2,t,{\bf p})$ can be Fourier transformed into a function defined on
3-dim cubic lattice in ${\bf k}$ space
which is reciprocal to the original lattice.
\begin{equation}
\tilde{D}(t,{\bf k})=\sum_{\bf p} D(1,2,t,{\bf p})e^{i{\bf kp}}.
\end{equation}
To incorperate the discreteness of the lattice we define a
mode spectrum as follows:
\begin{equation}
S(t,k)=k\sum_{\bf \hat k} \mid \tilde{D}(t,{\bf k})\mid ^{2}.
\end{equation}
which gives the relative importance of each
wave length component to the divergence.
In Fig.2.a and Fig.2.b, $S(0,k)$ and $S(t_{f},k)$
for a typical run
are plotted as a function of $k$.
$t_{f}$ is the time when saturation
is reached. We can see that from $t=0$ to
$t=t_{f}$ only the scale of $S$ changes. The shapes
in the two cases
are indistinguishable. Both of them are almost flat when
$k<4$
which indicates the randomness of the difference
between the two configurations and decrease at
larger $k$. So we conclude there is no evidence
of any preference of the divergence of some specific wavelength
components of the distance. The decreasing
in $k>4$ region is an artificial effect
on a finite cubic lattice, that is,
there are fewer lattice sites for large $k$ on a
finite cubic lattice than we
expect for an infinite one. Most of the big
fluctuations are simply due to combinatorical properties on the
discrete lattice, e.g. it is easier to form a 5 by the
sum of squares of three integers than to form a 4
so we have a peak at $k={\sqrt 5}$.
Considering these artificial effects,
we think the lattice might be too
small for a conclusive analysis of this kind.

After several runs with different initial $T_{0}$'s, we
get a relation between $h$ and $E$ shown in fig.3.
Here $E$ is the averaged energy on each plaquette. Using
the scaling property of the Hamiltonian in (1), we can get a
parameter invariant linear relation:
\begin{equation}
ha=\frac{4}{5}g^{2}(\frac{E}{N_{c}^{2}-1})a=\frac{1}{10}g^{2}Ea.
\end{equation}
The number is estimated from the line in fig.3.
We see the points fit nicely on the straight line
which goes through the
origin. In the classical limit, $h$ is independent on $a$
because $g$ does not run with $a$ in this limit.
The difference between the
two points for the lowest $T_{0}$
is an indication of the statistical uncertainty
of the value $h$.
The fluctuations are small because of the large
number of degrees of freedom involved.
The lack of points at low energy is caused by the rapid increase of
time needed to thermalize a system at low
temperature.
Also as we lower the temperature, quantum effects will become more
and more important so the
correctness of a classical calculation will become doubtful.
We postpone this problem to the future study.
To a good approximation, $E$ is related to the temperature as \cite{ambjorn}:
\begin{equation}
E=\frac{16}{3}T.
\end{equation}
Combine (10) and (11), we get
\begin{equation}
h=0.54g^{2}T
\end{equation}

It is a general result of classical non-linear dynamics
that the sum of
all positive Lyapunov exponents
describes the entropy growth rate,
which is roughly the inverse of the time scale
that the system approaches thermal equilibrium
(for example, see \cite{zaslavsky}). So the
largest Lyapunov exponent leads to an estimate
of the thermalization time $\tau_{s}=h^{-1}$.
If we insert the thermal coupling constant \cite{muller}
\begin{equation}
g^{2}(T)=\frac{16\pi^{2}}{11\ln(\pi T /\Lambda)^{2}}
\end{equation}
(with $\Lambda=200 $ MeV), we get an expression
of $\tau_{s}(T)$.
As an example to show the time scale,
at $T=400$ MeV, $\tau_{s}$ is roughly 0.24 fm/c.
So here we really see that the gluon system thermalizes
very fast.

Finally we want to have a short discussion on the gluon damping rate.
The motivation here is the seemingly numerical accident
that the value of $h$ is remarkably
similar to twice the damping rate of a gluon at rest \cite{braaten},
\begin{equation}
\gamma(\omega=0) =0.264 g^{2}T,
\end{equation}
in a thermal gluon system.
This is a quite surprising result because these two quantities
appear in totally different contexts and are calculated
with different methods. On one hand,
the damping rate is
the imaginary part of the self energy of a quasi-particle
in a thermal gluon system and is calculated by a sophisticated
effective quantum field theory, while the Lyapunov exponent
here is a classical dynamical
quantity describing the divergent property of two
classical trajectories.
But though we cannot establish a direct relation between these two
quantities in the moment, we think this similarity does not
arise without any reason.
First these two quantities, though very different from their
contexts, both describe
how fast a non-equilibrated
gluon system approaches thermal equilibrium.
The relevance of $h$ is clear from the above discussion.
The connection of $\gamma(\omega)$ appears clearly \cite{weldon} in
\begin{equation}
f(\omega,t)=\frac{1}{e^{\omega /T} - 1}
+c(\omega)e^{-2\gamma (\omega) t},
\end{equation}
where $f$ is time dependent distribution function,
the first term on the l.h.s is the equilibrated distribution.
We see $\gamma$ appears in the decaying term.
Second, we want to prove that
the gluon damping rate is basically
a quantity of semi-classical origin.
Since $\gamma(\omega)$ is a smooth function, it is
sufficient to prove the statement for a gluon at rest
for which case explicit quantum field calculation
is given in Ref.\cite{braaten}.
So in the following, we will concentrate only on the
gluon at rest.

In Ref.\cite{braaten}, the damping rate
is obtained from the soft part of the thermal
self energy loops of a gluon at rest (Fig.4.a).
Equivalently, it can also be obtained by calculating cross sections
from the diagrams resulting by cutting the self energy loops,
multiplying by the thermal distribution functions and
integrating over phase space \cite{weldon}.
Or physically, we can first calculate the decay rate
$\gamma_{d}$ and the production
rate $\gamma_{p}$ of a gluon at rest in a thermal gluon
system and then substract them to get $\gamma = \gamma_{d}-\gamma_{p}$.
{}From the cutting of the self energy loop,
we can see that the contributions to $\gamma$
come from Rutherford scattering (Fig.4.b)
and bremsstrahlung processes (Fig.4.c).
Both of them need only to be calculated on the
tree level to get the
leading order contributions to $\gamma$ (of the order of $g^{2}T$),
provided the effective Green's functions are used
for the soft vertices and soft propagators appearing in the
diagrams. This latter requirement is equivalent to resummation
\cite{braaten2}.

On the other hand, we know generally the result of a tree diagram
in a massless field theory
is of classical limit \cite{zuber}.
Now in our finite temperature
case quantum effects can come from two
directions. First the effective vertices and propagators
might be of quantum origin. Secondly we must use the quantum
thermal distribution function in order to get a finite result
of thermal contribution.

First we consider the effective Green's functions.
We want to prove that these functions are of semi-classical
origin in the
high temperature limit.
It is shown \cite{braaten2} that in the high temperature limit
the leading order modification to
a soft propogator or vertex
is obtained by inserting a hard thermal
loop (HTL) in the bare quantity (Fig.5.a and
Fig.5.c). The hard lines themselves
need not be modified if we only require leading order
solutions. The thermal contributions
correspond to diagrams with
a cut through one internal line of the hard thermal
loop at any place (put the corresponding
line on mass shell).  This can be seen in the
real time formalism of finite temperature field theory.
In this formalism, a thermal propagator is the sum of the vacuum
propagator and another term due to finite temperature
$D \sim n(\omega)\delta(p^{2}-m^{2})$, where $n(\omega)$
is the thermal distribution function and the $\delta$-function
puts the particle on mass shell.
The multiplication of the vacuum part of the propagators
around the loop just gives the vacuum contribution which
leads to renormalization. The thermal contribution
comes from terms each of which has a $\delta$-function
from the second term of a propagator.
The cutting procedure changes the
original loop diagram into a tree level
diagram (Fig.5.b and Fig.5.d)
with thermal distribution functions attached
to the two new external legs.
So we see that the modification to a soft
propagator or a vertex due to a hard thermal
loop is at tree level, i.e., it is of semi-classical origin.
The quantum effects come only from the Bose distribution
functions.
To suggest a possible method to
actually obtain these HTL's classically,
here we mention that the polarization effect (2-point HTL)
in a QED plasma was obtained long ago within classical
kinetic theory \cite{silin}. Recently
the kinetic formalism has been used to calculate
all hard thermal loops \cite{blaizot}.

Now in principle we can start from these effective quantities
and construct a classical perturbative method.
Then we can use it to calculate bremsstrahlung processes
and Rutherford scattering
and finally obtain the gluon damping rate to order $g^{2}T$.

We have proved that $\gamma(0)$ is a semi-classical
quantity. We now further show that it is really classical.
We have seen that the quantum effects come into $\gamma(0)$
only through a trivial way, namely, via the Bose distribution
function. If we want to
use the classical phase space
distribution $n=T/\omega$, which corresponds to the
high temperature limit of Bose distribution, in the calculation,
we must use a ultraviolet cutoff (lattice size $a$)
in order to avoid divergences.
Then all the HTL's may depend on $a$.
Now to see how $a$ comes into a quantity, we shall
express this quantity in terms
of classical variables: $g_{c}^{2}=4\pi\alpha_{s}/\hbar$,
$T$ and $a$. Here $g_{c}$ is so defined that $\hbar$
does not appear explicitly in classical Yang-Mills equations.
For example,
the plasma frequency will be $\omega_{p}=g_{c} \sqrt{T/a}$,
which diverges when $a$ goes to zero.
Though $\omega_{p}$ can be calculated semi-classically, it is not well
defined at strict classical limit.
On the other hand $\gamma(0) \sim g^{2}_{c} T$ is independent on
this cutoff $a$ in the leading order. This proves that to the leading
order $\gamma(0)$ is a purely classical quantity.


To conclude, we have shown that classical SU(3) gauge theory is
chaotic. This provides a classical explanation for why the gluon system
thermalizes rapidly.

$Acknowledgements$:
I am deeply indebted to B. M\"uller for the insightful
discussions throughout this work. I also
want to thank A. Trayanov for his help on the computational
aspect of this work and M. Thoma for the discussions on the problem
of gluon damping rate. This work has been supported in part
by the U.S. Department of Energy (Grant No. DE-FG05-90ER40592) and
by a computing grant from the North Carolina Supercomputing Center.

\newpage
{\noindent \Large \bf Figure Captions}
\bigskip
\begin{itemize}
\item[Fig.1] Divergence of two classical trajectories in
two different measures $D(t)$ (a) and $D^{E}(t)$ (b).
\item[Fig.2] Wave-length spectrum of the local distance at $t=0$(a) and
$t=t_{f}$(b).
\item[Fig.3] Lyapunov exponent as a function of energy.
\item[Fig.4] (a) Soft part of the self energy loop for a gluon at rest,
 (b) Rutherford scattering and (c) bremsstrahlung process resulting
from the cutting of (a). The solid lines are for hard gluons with momenta of
the order of $T$ and the
dotted lines are for soft ones of the order $gT$.
The blobs denote effective Green's functions resulting from inserting hard
thermal
loops in diagrams with soft external gluon lines.
\item[Fig.5] (a) HTL contribution to the self energy
of a soft gluon, (b) tree level diagram
from cutting of (a), (c) HTL contribution to the soft three gluon vertex,
(d) tree level diagram from cutting of (c). Lines have same meaning
as that in Fig.4.
\end{itemize}

\end{document}